\documentclass[a4paper]{jpconf}
\usepackage{graphicx}

\newcommand{\bea}{\begin{eqnarray}}
\newcommand{\eea}{\end{eqnarray}}
\newcommand{\be}{\begin{equation}}
\newcommand{\ee}{\end{equation}}

\begin{document}
\title{
Pomeron in the ${\mathcal N}=4$ SYM
at strong
couplings}

\author{A V Kotikov$^1$, L N Lipatov$^2$}

\address{$^1$ Bogoliubov Laboratory of Theoretical Physics,
Joint Institute for Nuclear Research,
141980 Dubna, Russia}
\address{$^2$ Theoretical Physics Department, 
Petersburg Nuclear Physics Institute,
188300, Gatchina,
Russia}

\ead{kotikov@theor.jinr.ru, lipatov@mail.desy.de}

\begin{abstract}
We show the result for the BFKL Pomeron intercept 
at ${\mathcal N}=4$ Supersymmetric
Yang-Mills model 
in the form of the inverse coupling expansion
$j_0=2-2\lambda^{-1/2}-\lambda^{-1}
+ 1/4 \, \lambda^{-3/2} + 2(1+3\zeta_3)\lambda^{-2}
+ O(\lambda^{-5/2})$, which has been calculated recently in 
\cite{Kotikov:2013xu} with the use of the AdS/CFT correspondence.

\end{abstract}

\section{Introduction}

The investigation of the high energy behavior of scattering amplitudes in the
${\mathcal N}=4$ Supersymmetric
Yang-Mills (SYM) model \cite{KL00}-\cite{Fadin:2009} is important for
our
understanding of the Regge processes in QCD. Indeed, this conformal model
can be considered as a simplified version of QCD, in which
the next-to-leading order (NLO) corrections \cite{next,next1} to the
Balitsky-Fadin-Kuraev-Lipatov (BFKL) equation \cite{BFKL}-\cite{BFKL4}
are comparatively simple and numerically
small. In the ${\mathcal N}=4$ SYM the equations for composite states of 
several reggeized gluons and for anomalous dimensions (AD) of quasi-partonic
operators turn out to be integrable at the leading logarithmic 
approximation \cite{L93,L93a,L97}.
Further, the eigenvalue of the BFKL kernel for this model
has the remarkable
property of
the maximal transcendentality
\cite{KL}. This property
gave a possibility to calculate the AD
$\gamma$ of the twist-2 Wilson operators in one
\cite{Lipatov00,Dolan:2001tt},
two \cite{KL,Kotikov:2003fb}, three \cite{Kotikov:2004er}, four
\cite{Kotikov:2007cy,Janik} and five \cite{Lukowski:2009ce} loops using
the QCD results \cite{Moch:2004pa}
and the asymptotic  Bethe ansatz \cite{Beisert:2005fw} improved with
wrapping corrections \cite{Janik}
in an agreement with the BFKL predictions  \cite{KL00,KL}.

On the other hand, due to the AdS/CFT-correspondence
\cite{AdS-CFT,AdS-CFT1,Witten}, in ${\mathcal N}=4$ SYM
some physical quantities can be also computed
at large couplings.
In particular, for AD of the large spin operators
Beisert, Eden and Staudacher constructed
the integral equation \cite{Beisert:2006ez} with the use
the asymptotic
Bethe-ansatz. This equation reproduced the known results
at small coupling constants
and
is in a full agreement (see \cite{Benna:2006nd,Kotikov:2006ts,Basso:2007wd}) 
with large
coupling predictions
\cite{Gubser:2002tv,Frolov:2002av,Roiban:2007jf}.

With the use of the BFKL equation in a diffusion approximation
\cite{KL00,Fadin:2007xy},
strong coupling results
for AD
\cite{Gubser:2002tv,Frolov:2002av,Roiban:2007jf} 
and the pomeron-graviton duality 
\cite{Polchinski:2001tt,Polchinski:2002jw}
the Pomeron intercept was calculated
at the leading order in the inverse
coupling constant (see the Erratum\cite{Kotikov:2006}
to the paper \cite{Kotikov:2004er}).
Similar results 
were obtained also in
Ref. \cite{Brower:2006ea}.
The Pomeron-graviton duality in the   ${\mathcal N}=4$ SYM
gives a possibility to construct the Pomeron interaction model
as a generally covariant effective theory for the reggeized gravitons
\cite{Lipatov:2011ab}.

Below we present the strong coupling corrections
to the Pomeron intercept $j_0=2-\Delta$ in next orders. These corrections
were obtained in Ref. \cite{Kotikov:2013xu} 
 with the use of the recent
calculations \cite{Gromov:2011de}-\cite{Roiban:2011fe}
of string energies.

\section{BFKL  equation at small coupling constant}

The eigenvalue of the BFKL equation in  ${\mathcal N}=4$ SYM model has
the following perturbative expansion
\cite{KL00,KL}
(see also Ref.
\cite{Fadin:2007xy})
\be
j-1 = \omega = \frac{\lambda}{4\pi^2} \biggl[\chi(\gamma_{BFKL}) +
\delta(\gamma_{BFKL})
\frac{\lambda}{16\pi^2}\biggr], ~~~ \lambda=g^2N_c,
\label{1i}
\ee
where $\lambda$ is the t'Hooft coupling constant.
The quantities $\chi$ and $\delta$
are functions of
the conformal weights
$m$ and $\widetilde{m}$ of the principal series of unitary M\"{o}bius group representations, but for the
conformal spin $n=m-\widetilde{m}=0$ they depend only on the BFKL anomalous dimension
\be
\gamma_{BFKL}=\frac{m+\widetilde{m}}{2}=\frac{1}{2}+i\nu \,
\label{2i}
\ee
and are presented below \cite{KL00,KL}
\bea
\chi(\gamma) &=& 2\Psi(1)-\Psi(\gamma)-\Psi(1-\gamma), \label{2i} \\
\delta(\gamma) &=& \Psi^{''}(\gamma)+\Psi^{''}(1-\gamma) + 6\zeta_3
-2\zeta_2 \chi(\gamma) -2\Phi(\gamma)-2\Phi(1-\gamma) \, .
\label{3i}
\eea

Here $\Psi (z)$ and $\Psi ^{\prime }(z)$, $\Psi ^{\prime \prime }(z)$ are
the Euler $\Psi $ -function and its derivatives.
The function $\Phi(\gamma)$ is defined as follows
\be
\Phi(\gamma) = 2 \sum_{k=0}^{\infty }
\frac{1}{k+\gamma} \, \beta ^{\prime }(k+1)\,,~~~
\beta ^{\prime }(z)=\frac{1}{4}\Biggl[
\Psi ^{\prime }\Bigl(\frac{z+1}{2}\Bigr)-
\Psi ^{\prime }\Bigl(\frac{z}{2}\Bigr)\Biggr]\,.
\label{5i}
\ee

Due to the symmetry of $\omega$ to the substitution
$\gamma _{BFKL}\rightarrow 1-\gamma _{BFKL}$ expression (\ref{1i}) is an
even function of $\nu$
\be
\omega = \omega_0 + \sum_{m=1}^{\infty} (-1)^m \, D_m \, \nu^{2m} \, ,
\label{6i}
\ee
where
\bea
 \omega_0 &=& 4\ln 2 \, \frac{\lambda}{4\pi^2} \left[ 1- \overline{c}_1
\frac{\lambda}{16\pi^2}  \right] +
O(\lambda^3) \,, \label{7i} \\
D_m &=&
2\left(2^{2m+1}-1\right)\zeta_{2m+1} \frac{\lambda}{4\pi^2} +\frac{\delta ^{(2m)}(1/2)}{(2m)!}\,
\frac{\lambda ^2}{64 \pi ^4}+
O(\lambda^3) \,.
\label{8i}
\eea
According to Ref. \cite{KL} we have
\be
 \overline{c}_1 ~=~ 2 \zeta_2 +
\frac{1}{2\ln 2} \left(11\zeta_3
- 32{\rm Ls}_{3}\Bigl(\frac{\pi }{2}\Bigl) -14 \pi \zeta_2
\right) \approx  7.5812 \,, ~~~
{\rm Ls}_{3}(x)=-\int_{0}^{x}\ln ^{2}\left| 2\sin \Bigl(\frac{y}{2}%
\Bigr)\right| dy \,. \label{8i}
\ee

Due to the M\"{o}bius invariance and hermicity of the BFKL hamiltonian in ${\mathcal N}=4$ SYM
expansion (\ref{6i})
is valid also at large coupling constants.
In the framework of the AdS/CFT correspondence the BFKL Pomeron is equivalent to
the reggeized graviton
~\cite{Polchinski:2002jw}.
In particular,
in the strong coupling regime $\lambda \rightarrow \infty$
\begin{equation}
j_0~=~ 2-\Delta \,,
\label{11i}
\end{equation}
where the leading contribution
$\Delta =
2/\sqrt{\lambda}$ was
calculated in Refs.~\cite{Kotikov:2006, Brower:2006ea}. Below we find NLO
terms in
the strong coupling expansion
of the Pomeron intercept.

\section{ AdS/CFT correspondence}

Due to the energy-momentum conservation, the universal AD
of the stress tensor $T_{\mu \nu}$ should be zero, i.e.,
\be
\gamma(j=2)=0.
\label{1e}
\ee

It is important, that the AD
$\gamma$ contributing to the DGLAP
equation \cite{DGLAP}-\cite{DGLAP3}
does not coincide with $\gamma_{BFKL}$ appearing in the BFKL equation.
They are related as follows \cite{next,Salam:1998tj,Salam:1999cn} 
\be
\gamma ~=~ \gamma_{BFKL} + \frac{\omega}{2} ~=~ \frac{j}{2}+i\nu \,,
\label{2e}
\ee
where the additional contribution $\omega /2$ is responsible in particular for the cancelation
of the singular terms
$\sim 1/\gamma ^3$  obtained from the NLO
corrections (\ref{1i}) to the eigenvalue of the BFKL kernel
\cite{next}.
Using  above relations
one obtains
\be
\nu(j=2)=i\,.
\label{4e}
\ee
As a result, from eq. (\ref{6i}) for the Pomeron trajectory
we derive the following
representation for the correction
$\Delta$  (\ref{11i})
to the graviton spin $2$
\be
\Delta ~=~  \sum_{m=1}^{\infty} D_m .
\label{5e}
\ee

According to
(\ref{11i}) and (\ref{5e}), we have the following
small-$\nu$ expansion for the eigenvalue of the BFKL kernel
\be
j-2 = \sum_{m=1}^{\infty}  D_m \left({(-\nu^{2})}^m-1\right),
\label{7e}
\ee
where $\nu^2$ is related to $\gamma$ according to eq. (\ref{2e})
\be
\nu^2=
-{\left(\frac{j}{2}
-\gamma\right)}^2.
\label{8e}
\ee

On the other hand, due to the
 ADS/CFT correspondence the string energies $E$ in dimensionless units are related to the AD
$\gamma$ of the twist-two
operators as follows
\cite{AdS-CFT1,Witten}\footnote{Note
that our expression (\ref{1a}) for the  string energy $E$ differs from a
definition, in which
$E$ is equal to the scaling dimension $\Delta_{sc}$. But eq.
(\ref{1a}) is correct, because it can be presented as $E^2=(\Delta_{sc}-2)^2-4$ and
coincides with Eqs. (45) and (3.44) from Refs. \cite{AdS-CFT1} and
\cite{Witten}, respectively.}

\begin{equation}
E^2=(j+\Gamma)^2-4,~~\Gamma=-2\gamma
\label{1a}
\end{equation}
and therefore we can obtain from (\ref{8e}) the relation between the parameter $\nu$ for
the principal series of unitary representations of the M\"{o}bius group and the string
energy $E$
\be
\nu^2=
-\left(\frac{E^2}{4} +1\right)\,.
\label{3a}
\ee
This expression for $\nu ^2$ can
be inserted in the r.h.s. of Eq. (\ref{7e}) leading to the following expression for the Regge trajectory
of the graviton in the anti-de-Sitter space
\be
j-2 =  \sum_{m=1}^{\infty}
D_m \left[{\left(\frac{E^2}{4}+1\right)}^m-1\right].
\label{4a}
\ee

\section{Graviton Regge trajectory and Pomeron intercept}

We
assume,
that eq. (\ref{4a})  is valid also
at large $j$ and large $\lambda$ in the region
$1\ll j \ll\sqrt{\lambda}$,
where the strong coupling calculations of energies
were performed~\cite{Gromov:2011de,Roiban:2011fe}.
These energies
can be presented  in the form
\footnote{Here we put $S=j-2$, which in particular is related to the
use of the angular momentum $J_{an}=2$ in calculations of Refs \cite{Gromov:2011de,Roiban:2011fe}.}
\be
\frac{E^2}{4} ~=~ \sqrt{\lambda} \, \frac{S}{2}\, \left[h_0(\lambda)
+ h_1(\lambda) \frac{S}{\sqrt{\lambda}} + h_2(\lambda) \frac{S^2}{\lambda}
\right] + O\Bigl(S^{7/2}\Bigr),
\label{5a}
\ee
where
\be
 h_i(\lambda) ~=~  a_{i0} + \frac{a_{i1}}{\sqrt{\lambda}} +
\frac{a_{i2}}{\lambda} +  \frac{a_{i3}}{\sqrt{\lambda^3}} +
\frac{a_{i2}}{\lambda^2}.
\label{5.1a}
\ee

 The contribution $\sim \sqrt{S}$ can be extracted directly from
the
Basso result \cite{Basso:2011rs,Basso:2012ex}
taking $J_{an}=2$ according to \cite{Gromov:2011bz}:
\be
h_0(\lambda) =
\frac{I_3(\sqrt{\lambda})}{I_2(\sqrt{\lambda})} + \frac{2}{\sqrt{\lambda}} =
\frac{I_1(\sqrt{\lambda})}{I_2(\sqrt{\lambda})} - \frac{2}{\sqrt{\lambda}}\, ,
\label{Ad5.1}
\ee
where $I_k(\sqrt{\lambda})$ is the modified Bessel functions.
It leads to the following values of  coefficients $a_{0i}$
\be
a_{00} ~=~ 1,~~ a_{01}~=~ - \frac{1}{2},~~a_{02} ~=~ a_{03}~=~  \frac{15}{8},~~
a_{04}~=~  \frac{135}{128}
\label{Ad5.2}
\ee

The coefficients $a_{10}$ and $a_{20}$ come from considerations of the
classical part of
the folded spinning string corresponding to the twist-two
operators 
(see, for example, \cite{Roiban:2011fe})
\be
a_{10}~=~  \frac{3}{4},~~a_{20} ~=~ - \frac{3}{16}\,.
\label{Ad5.3}
\ee

The one-loop coefficient $a_{11}$ is found recently in
the paper
 \cite{Gromov:2011bz},
considering different
asymptotical regimes with taking into account the Basso result 
\cite{Basso:2011rs} ($\zeta_3$ is the Euler $\zeta$-function)
\be
a_{11}~=~ \frac{3}{16} \Bigl(1-\zeta_3\Bigr),
\label{Ad5.4}
\ee

Comparing the l.h.s. and r.h.s. of (\ref{4a}) at large $j$ values
gives us the
coefficients $D_m$ and $\Delta$ (see Appendix A in \cite{Kotikov:2013xu}).

\section{ Conclusion}

We have shown the intercept of the BFKL pomeron at weak coupling regime and
demonstrated an approach to obtain its  values at strong couplings
(for details, see Ref. \cite{Kotikov:2013xu}).

At 
$\lambda \rightarrow \infty$, the correction $\Delta$ for the
Pomeron
intercept $j_0=2-\Delta$ has the form 
\footnote{Using a similar approach, the coefficients $\sim \lambda^{-1}$ and 
$\sim \lambda^{-3/2}$
were calculated also in the paper \cite{Costa:2012cb}.
After correction of some errors, the results in \cite{Costa:2012cb} coincide 
with ours.}
\be
\Delta ~=~  \frac{2}{\lambda^{1/2}} \, \left[1
+
\frac{1}{2\lambda^{1/2}} - \frac{1}{8\lambda} -
\Bigl(1 +3 \zeta_3 \Bigr) \frac{1}{\lambda^{3/2}} +
\left(2a_{12} - \frac{145}{128} - \frac{9}{2}\zeta_3 \right)
\frac{1}{\lambda^2} + O\left(\frac{1}{\lambda^{5/2}}\right)
\right] \, . \label{11.de}
\ee

The fourth corrections in (\ref{11.de})
contain unknown coefficient $a_{12}$,
which will be obtained after
the evaluation of spinning folded string on the two-loop level.
Some estimations were given in Section 6 of \cite{Kotikov:2013xu} .

\ack
A.V.K. was supported in part
by RFBR grant No. 13-02-01060-a.
He thanks Eugene Levin and the Organizing Committee of 
International Moscow Phenomenology Workshop 
for invitation.

\end{document}